\documentclass[a4paper,11pt]{article}
\usepackage{jheppub}
\usepackage{lineno}

\usepackage{xspace}
\newcommand{\msbar}{\ensuremath{\mathrm{\overline{MS}}}\xspace}
\newcommand{\ttbar}{\ensuremath{\mathrm{t\overline{t}}}\xspace}
\newcommand{\mtt}{\ensuremath{m_\ttbar}\xspace}

\newcommand{\mt}{\ensuremath{m_\mathrm{t}}\xspace}
\newcommand{\mtmt}{\ensuremath{\mt(\mt)}\xspace}
\newcommand{\mur}{\ensuremath{\mu_\mathrm{r}}\xspace}
\newcommand{\muf}{\ensuremath{\mu_\mathrm{f}}\xspace}
\newcommand{\mum}{\ensuremath{\mu_\mathrm{m}}\xspace}
\newcommand{\muk}{\ensuremath{\mu_k}\xspace}
\newcommand{\mtmu}{\ensuremath{\mt(\mum)}\xspace}
\newcommand{\mtmuk}{\ensuremath{\mt(\muk)}\xspace}
\newcommand{\mtmukk}{\ensuremath{\mt(\muk/2)}\xspace}
\newcommand{\chisq}{\ensuremath{\chi^2}\xspace}
\newcommand{\as}{\ensuremath{\alpha_\mathrm{S}}\xspace}
\newcommand{\asmz}{\ensuremath{\as(m_\mathrm{Z})}\xspace}
\newcommand{\Matrix}{\textsc{Matrix}\xspace}
\newcommand{\ABMP}{ABMP16\_5\_nnlo\xspace}
\newcommand{\ie}{i.e.\ }
\newcommand{\eg}{e.g.\ }

\arxivnumber{2208.11399}

\title{\boldmath Running of the top quark mass at NNLO in QCD}

\author[a,1]{Matteo M. Defranchis,\note{Corresponding author.}}
\author[a,b]{Jan Kieseler,}
\author[c,d]{Katerina Lipka,}
\author[e,f]{Javier Mazzitelli}
\affiliation[a]{CERN, Geneva, Switzerland}
\affiliation[b]{Karlsruhe Institute of Technology, Karlsruhe, Germany}
\affiliation[c]{Deutsches Elektronen-Synchrotron DESY, Hamburg, Germany}
\affiliation[d]{University of Wuppertal, Germany}
\affiliation[e]{Max Planck Institute for Physics, Munich, Germany}
\affiliation[f]{Paul Scherrer Institut, Villigen PSI, Switzerland}

\emailAdd{matteo.defranchis@cern.ch}
\emailAdd{jan.kieseler@cern.ch}
\emailAdd{katerina.lipka@desy.de}
\emailAdd{javier.mazzitelli@psi.ch}

\abstract{The running of the top quark mass (\mt) is probed at the next-to-next-to-leading order in quantum chromodynamics for the first time. The result is obtained by comparing calculations in the modified minimal subtraction (\msbar) renormalisation scheme to the CMS result on differential measurement of the top quark-antiquark (\ttbar) production cross section at $\sqrt{s} = 13~\mathrm{TeV}$.
The scale dependence of \mt is extracted as a function of the invariant mass of the \ttbar system, up to an energy scale of about 0.5~TeV. The observed running is found to be in good agreement with the three-loop solution of the renormalisation group equations of quantum chromodynamics.}

\begin{document}
\maketitle
\flushbottom

\section{Introduction}
\label{sec:intro}

In the modified minimal subtraction (\msbar) renormalisation scheme, the parameters of the quantum chromodynamics (QCD) lagrangian, \ie the strong coupling constant \as and the masses of the quarks, depend on the energy scale at which they are evaluated. This effect, often referred to as ``running'', is described by the renormalisation group equations (RGEs) of QCD, which can be solved using perturbation theory. The running of the quark masses has been calculated up to order $\as^5$~\cite{Baikov:2014qja,Luthe:2016xec}.
Measurements of the running of quark masses are not only a proof of QCD as a renormalisable theory, but also an indirect probe of physics beyond the standard model. In fact, the QCD RGE would be modified \eg in the context of supersymmetric theories~\cite{Mihaila:2013wma} or in models implying dynamic mass generation~\cite{Christensen:2005hm}.

Experimentally, the running of the charm quark mass was investigated using deep inelastic scattering data at the DESY HERA~\cite{Gizhko:2017fiu}, while the running of the bottom quark mass has been demonstrated using results from the CERN LEP, SLAC SLC, DESY HERA, and CERN LHC~\cite{Behnke:2015qja,Aparisi:2021tym}, up to the scale of the Higgs boson mass. The running of the top quark mass has been investigated for the first time by the CMS Collaboration at the CERN LHC~\cite{CMS:2019jul}. The \mt running was extracted from a measurement of the \ttbar production cross section as a function of the invariant mass of the \ttbar system, \mtt, 
at $\sqrt{s} = 13~\mathrm{TeV}$. The measurement is used together with the QCD calculation at next-to-leading order (NLO) in the \msbar scheme~\cite{Dowling:2013baa} implemented in the MCFM program~\cite{Campbell:2010ff,Campbell:2012uf}, which was the state-of-the art theory to that date.

In this work, the published results of the CMS analysis~\cite{CMS:2019jul} are reinterpreted by using dedicated theory developments~\cite{Catani:2020tko}, bringing the measurement of the running of the top quark mass to the NNLO level in QCD, for the first time. Following the methodology of Ref.~\cite{Catani:2020tko}, the \mtt distribution obtained at NNLO with the \Matrix framework~\cite{Grazzini:2017mhc} in the pole mass scheme are translated into the \msbar scheme, and used together with the CMS measurement of Ref.~\cite{CMS:2019jul} which was designed to maximise the sensitivity to the running of \mt considering the experimental resolution.
The result presented in this paper benefits from an improved fit procedure, which allows for a consistent treatment of the numerical uncertainty in the theoretical predictions\footnote{The numerical uncertainties in our NNLO predictions include a statistical component from the Monte Carlo integration, as well as the $q_T \to 0$ extrapolation uncertainties that are intrinsic to the $q_T$-subtraction method~\cite{Catani:2007vq}. For more details, see Ref.~\cite{Grazzini:2017mhc}.}. This becomes necessary due to the increased numerical uncertainties in the NNLO calculations, limited by computing time and resources. This is however a small effect compared to the significantly reduced scale uncertainties in the NNLO prediction, which are at least a factor 2 smaller compared to the corresponding NLO calculation~\cite{Catani:2020tko}. In this work, the variations of the renormalisation and factorisation scales are fully taken into account.

\section{Theoretical setup and experimental inputs}
\label{sec:inputs}

In the CMS analysis of Ref.~\cite{CMS:2019jul}, the dependence of the running top quark mass \mtmu is investigated as a function of the scale $\mum = \mtt$, where \mtt is the invariant mass of the \ttbar system. 
In the calculation, the renormalisation (\mur), factorisation (\muf), and top quark mass (\mum) scales are all set to the value of \mt. In each bin of \mtt independently, the value of \mtmt is extracted by performing a \chisq fit of the theoretical calculation to the measured cross section. The extracted values of \mtmt are then converted to the corresponding values of \mtmuk using one-loop solutions of the RGEs, where \muk is the representative energy scale of bin~$k$ in \mtt, corresponding to the average \mtt value in that bin. The bin boundaries for \mtt and the corresponding values of \muk are reported in Table~\ref{tab:bins}.

\begin{table}[htb]

    \centering
    \begin{tabular}{c|cc}
    \hline
        bin ($k$) & \mtt [GeV] & \muk [GeV] \\ \hline
        1 & $<420$ & 384 \\
        2 & $420-550$ & 476 \\
        3 & $550-810$ & 644 \\
        4 & $> 810$ & 1024 \\
        \hline
    \end{tabular}
    \caption{Boundaries of the \mtt bins and representative energy scale for each bin $k$ of \mtt as defined in Ref.~\cite{CMS:2019jul}. The scales \muk are defined as the average value of \mtt in the corresponding bin.}
\label{tab:bins}
\end{table}

Following the approach suggested in Ref.~\cite{Catani:2020tko}, the CMS analysis was repeated by setting the scale \mum to $\muk/2$, independently in each bin of \mtt. This choice is preferred over \muk due to the fact that $\muk/2$ corresponds approximately to \mt in the vicinity of the \ttbar production threshold, which is the value typically used in the calculation of the total cross section. 
Furthermore, the bin-by-bin dynamic scale choice allows the value of \mtmukk to be determined directly. The two approaches were found to yield consistent results~\cite{Defranchis:2021eos}.

In this work, the approach proposed in Ref.~\cite{Catani:2020tko} is adopted, and NNLO calculations for \mtt distribution are used for the extraction of \mtmu. Unlike in Ref.~\cite{Defranchis:2021eos}, the \mur, \muf, and \mum  scales in the calculation are set to $\muk/2$, and scale uncertainties are estimated by varying \mur and \muf by a factor of two, avoiding cases in which $\mur/\muf = 4$ or $1/4$. The scale \mum is not varied in this context, as it represents the variable with respect to which the running is extracted. The calculation is interfaced with the \ABMP~\cite{Alekhin:2017kpj} set of parton distribution functions (PDFs), and the \ttbar production cross section is calculated in each bin of \mtt for different values of \mtmukk. In the calculation, the value of \asmz is set to 0.1147,  consistently with the utilised PDF set~\cite{Alekhin:2017kpj}. A comparison between the NLO and NNLO predictions and the CMS measurement of Ref.~\cite{CMS:2019jul} can be found in Ref.~\cite{Catani:2020tko}. The PDF uncertainties are estimated by performing the calculation using the complete set of PDF eigenvectors. In each bin, the PDF uncertainties are estimated with respect to a reference mass point, chosen such that the calculated cross section for that particular value of \mtmukk is close to the measured one in order to minimise any possible extrapolation bias. The relative PDF uncertainties obtained in this way are assumed to be independent of the value of the \mtmukk. This approximation is necessary in order to keep the computing time within an acceptable range, and its validity was verified using the corresponding NLO calculations.  Finally, in the case of the \ABMP PDFs, the uncertainty in the value of \as is included in the PDF variations.

\begin{figure}[htbp]
    \centering
    \includegraphics[width=.495\textwidth]{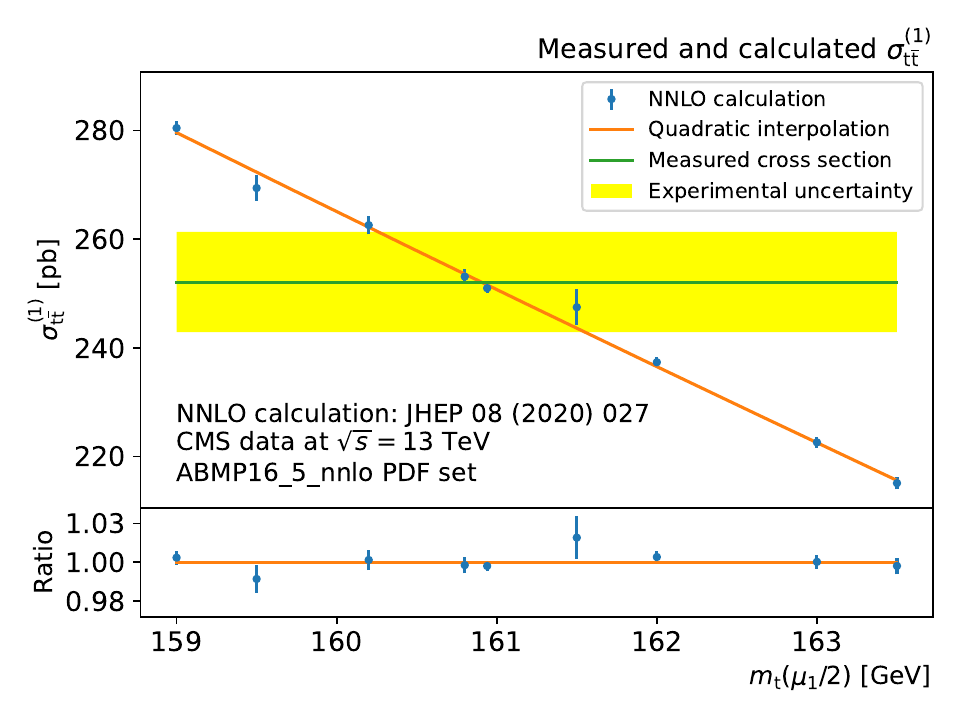}
    \includegraphics[width=.495\textwidth]{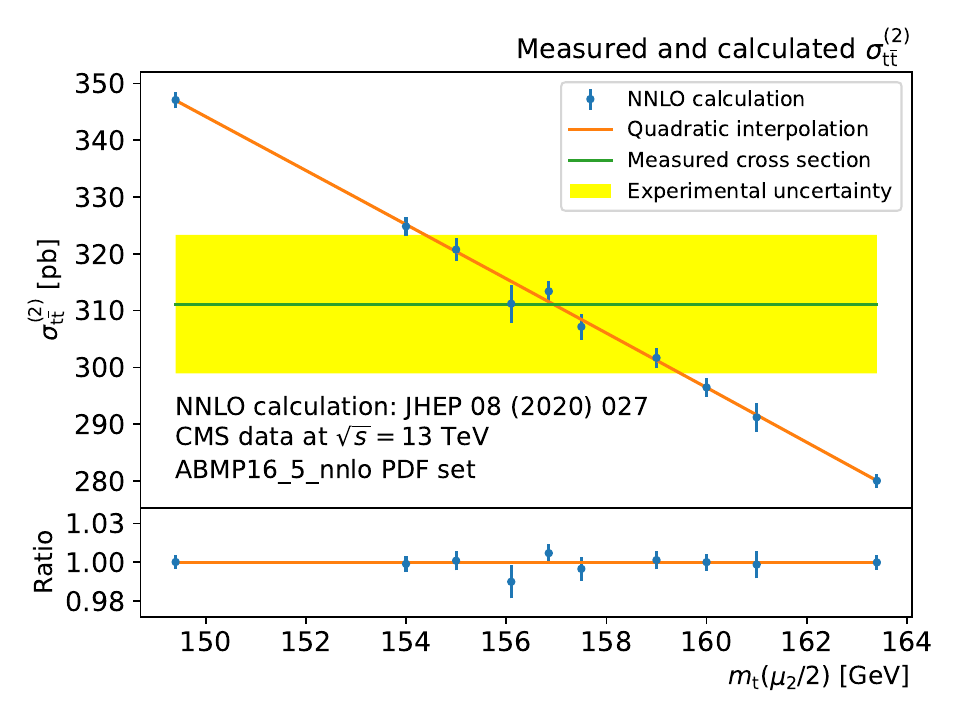}\\
    \includegraphics[width=.495\textwidth]{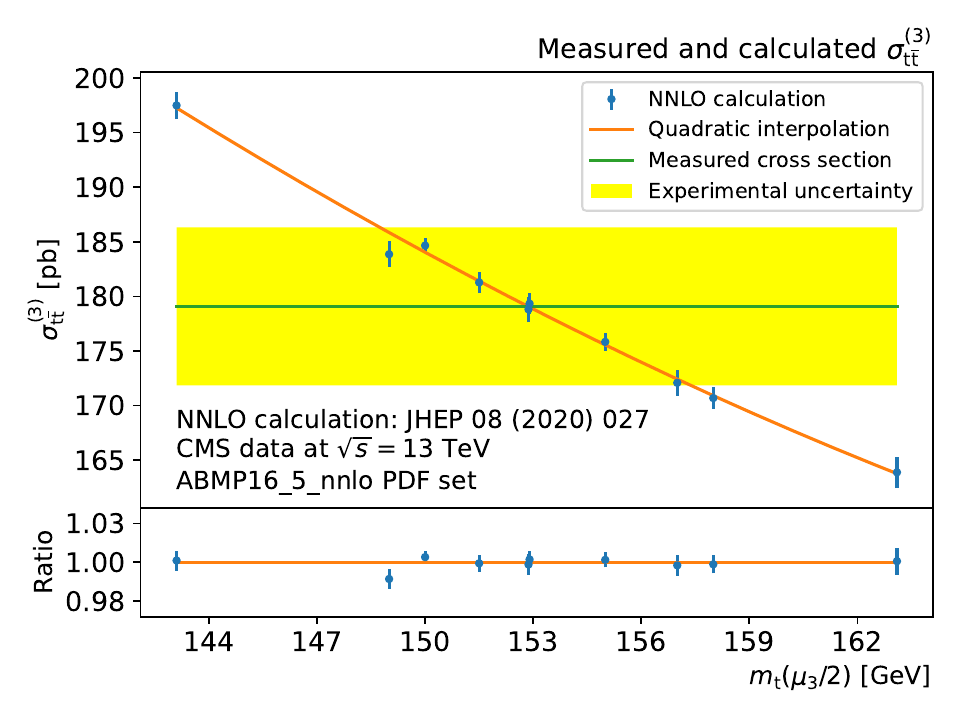}
    \includegraphics[width=.495\textwidth]{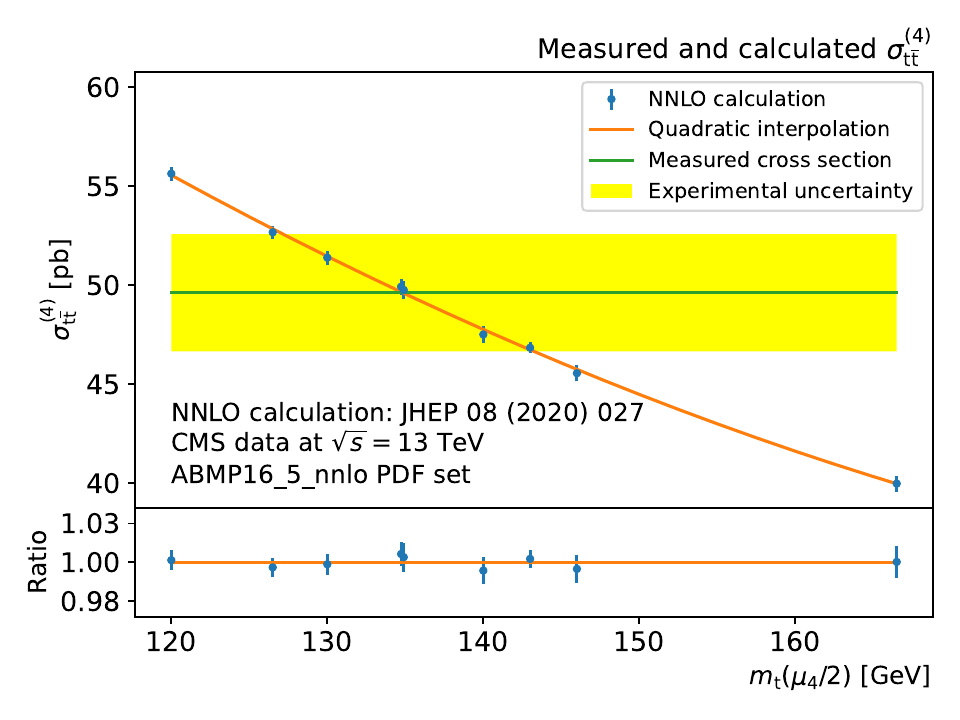}
    \caption{Calculated \ttbar production cross section $\sigma_\ttbar^{(k)}$ in bin $k$ of \mtt as a function of \mtmukk (points) compared to the re-scaled value of the measured cross section of Ref.~\cite{CMS:2019jul} (horizontal lines). The vertical error bars represent the numerical uncertainty in the theoretical predictions, while the horizontal error bands correspond to the re-scaled uncertainty in the measured cross sections. The dependence of the calculated cross section on the value of \mtmukk is parameterised assuming a quadratic dependence (line). The lower panels show the ratio between the calculated cross sections and the corresponding quadratic interpolations.}
    \label{fig:xsec}
\end{figure}

The differential cross section measured in Ref.~\cite{CMS:2019jul} is re-scaled in order to take into account the latest measurement of the total integrated luminosity by the CMS experiment~\cite{CMS:2021xjt}, which in the CMS analysis is one of the leading uncertainties in the extracted running~\cite{CMS:2019jul,Defranchis:2020efl}. This implies both a shift in the central values of the measured differential cross section and a reduction in the luminosity uncertainty from 2.5\% to 1.2\%. The covariance matrix between the bins is re-calculated accordingly. The obtained results are compared to the NNLO theoretical predictions as a function of \mtmukk in Figure~\ref{fig:xsec}. The dependence of the calculated cross section on the values of \mtmukk is found to be well described by a second-order polynomial.

\section{Fit procedure}
\label{sec:fit}

In order to properly take into account the correlations between the numerical uncertainties in the theoretical predictions and the PDF variations, an improved fit procedure compared to the one of Ref.~\cite{CMS:2019jul} has been developed. The values of \mtmukk are extracted simultaneously by means of a \chisq fit of the theoretical prediction to the measured differential cross section. The \chisq is parameterised as a function of the values of \mtmukk and of the nuisance parameters representing the effect of the numerical uncertainties in the NNLO calculation and the PDF uncertainties in the predicted differential cross section. This allows the numerical uncertainties and their correlations with the other parameters of the fit to be consistently taken into account, avoiding any possible bias in the determination of the running. The PDF uncertainties are estimated according to the \ABMP prescription~\cite{Alekhin:2017kpj}. The \chisq function is defined as:
\begin{equation}
    \chisq (\vec{m},\vec{j},\vec{\eta}) = \vec{\Delta}^\mathrm{T} (\vec{m},\vec{j},\vec{\eta}) \, C_\mathrm{exp}^{-1} \, \vec{\Delta} (\vec{m},\vec{j},\vec{\eta}) + \sum_{p=1}^\text{nPDF}j_p^2 + \sum_{t=1}^\text{nPred} \eta_t^2 \, ,
    \label{eq:chisq}
\end{equation}
where
\begin{equation}
    \Delta_k (m_k,\vec{j},\vec{\eta}) = \sigma_\mathrm{exp}^k-\sigma_\mathrm{th}^k(m_k,\vec{j},\vec{\eta})\, .
    \label{eq:residual}
\end{equation}

Here, $\vec{m}$ represents the set of free parameters used to determine the values of \mtmukk, while $\vec{j}$ and $\vec{\eta}$ are the nuisance parameters modelling the effect of the PDF uncertainties and numerical uncertainties in the calculated cross sections, respectively. Furthermore, $\sigma_\mathrm{exp}^k$ and $\sigma_\mathrm{th}^k$ correspond to the  values of the measured and calculated cross sections in bin $k$ of \mtt, respectively, the latter depending on the $\vec{m}$, $\vec{j}$, and $\vec{\eta}$ parameters. The matrix $C_\mathrm{exp}$ represents the covariance between the bins of the measured differential cross section, and it includes the effect of the experimental and extrapolation uncertainties described in Ref.~\cite{CMS:2019jul}. For asymmetric extrapolation uncertainties, the maximum between the positive-side and negative-side variations is conservatively taken. The first term in Eq.~\ref{eq:chisq}  is the statistical term, while the two following ones are Gaussian penalty terms representing the prior assumptions on the nuisance parameters. In the fit, all nuisance parameters are defined such that they follow a standard normal distribution. The index $p$ runs up to the number of PDF variations, $\mathrm{nPDF} = 29$, while the index $t$ runs up to the number of theoretical predictions used in the fit, $\mathrm{nPred}$, which include the 38 mass points and the 29 PDF variations.

The effect of the numerical uncertainties is modelled by introducing modifiers to the calculated cross sections that depend on the corresponding nuisance parameter and the size of the numerical uncertainty. For each calculated cross section $\sigma_\mathrm{th}^t$, including those obtained for the various PDF eigenvectors, the quantity $\sigma_\mathrm{th}^t(\eta_t)$ is defined:

\begin{equation}
    \sigma_\mathrm{th}^t(\eta_t) = \sigma_\mathrm{th}^t(1 + \eta_t \delta^t_\mathrm{num})\, ,
    \label{eq:num_unc}
\end{equation}
where $\delta^t_\mathrm{num}$ is the relative numerical uncertainty in $\sigma_\mathrm{th}^t$. For each nominal mass point $m$ in bin $k$ of \mtt, the dependence of the calculated cross section on the PDF variations is then estimated as:
\begin{equation}
    \sigma_\mathrm{th}^{m,k}(\vec{j},\vec{\eta}) = \sigma_\mathrm{th}^{m,k}(\eta_{m,k}) \prod_{p=1}^{\mathrm{nPDF}}\left[1 + j_p \left(\frac{\sigma_\mathrm{th}^{p,k}(\eta_{p,k})}{\sigma_\mathrm{th}^{m_0,k}(\eta_{m_0,k})}-1\right)\right] \, ,
    \label{eq:pdf}
\end{equation}
where $\sigma_\mathrm{th}^{m_0,k}$ is the nominal cross section for the reference mass point $m_0$ used to derive the PDF variations in bin $k$ (see Section~\ref{sec:inputs}), and $\sigma_\mathrm{th}^{p,k}$ is the calculated cross section corresponding to the PDF variation $p$ for the reference mass point. In Eq.~\ref{eq:pdf}, all calculated cross section are corrected for their numerical uncertainties according to Eq.~\ref{eq:num_unc}.

The quantities $\sigma_\mathrm{th}^{m,k}(\vec{j},\vec{\eta})$ are then used to derive the dependence of the calculated cross section on $m_k = \mtmukk$. For each choice of values for $\vec{j}$ and $\vec{\eta}$, the dependence $\sigma^k_\mathrm{th}(m_k)$ is estimated by means of a quadratic interpolation, as shown in Figure~\ref{fig:xsec}.
This way, the theoretical dependence on $m_k$ is smoothed and the impact of the numerical uncertainties is mitigated.
Furthermore, the correlations between the different mass points introduced by the PDF variations are fully taken into account in the interpolation procedure. These correlations arise from the fact that a single mass point is used to derive the dependence on the PDF variations, as shown explicitly in Eq.~\ref{eq:pdf}.

Finally, the uncertainties related to the choice of \mur and \muf are estimated by repeating the fit for each of the 7-point scale variations described in Section~\ref{sec:inputs}. The maximum variation observed in each bin, which in all cases correspond to one of the combined variations of \mur and \muf, is conservatively taken as the scale uncertainty in that bin. The correlations between the scale variations in the different bins are kept track of, and an additional covariance matrix is derived.

\section{Results}
\label{sec:results}

\begin{figure}[htbp]
    \centering
    \includegraphics[width=.75\textwidth]{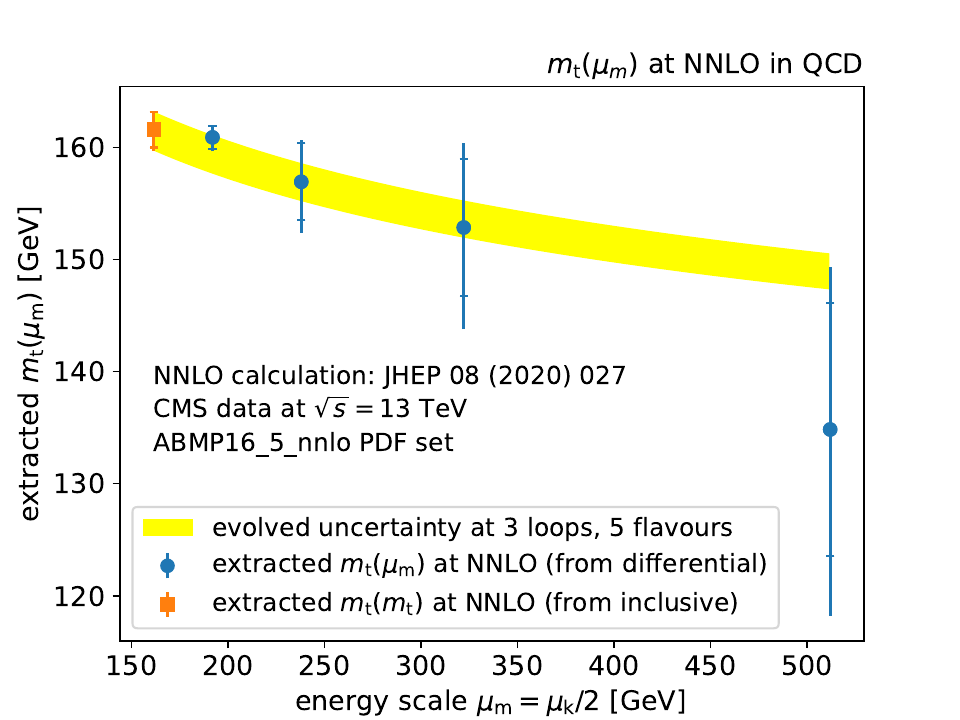}
    \caption{Extracted \mtmukk (circles) compared to the value of \mtmt (squared) obtained from the inclusive \ttbar production cross section~\cite{CMS:2018fks}. The inner vertical bars represent the combination of experimental, PDF, and numerical uncertainties, while the outer bars also include the QCD scale uncertainties. The band represent the evolved total uncertainty in \mtmt.}
    \label{fig:masses}
\end{figure}

In Figure~\ref{fig:masses}, the extracted \mtmukk are compared with the evolved value of \mtmt obtained in Ref.~\cite{CMS:2018fks}. The value of \mtmt was extracted from a measurement of the inclusive \ttbar cross section at $\sqrt{s} = 13~\mathrm{TeV}$ using NNLO predictions and the same PDF set as in this work. The numerical values of the \mtmukk are reported in Table~\ref{tab:masses}. The experimental (exp) uncertainty, corresponding to the total uncertainty in the measured differential cross section, is obtained by fixing all the $\vec{j}$ and $\vec{\eta}$ parameters to their post-fit values. The combination between PDF and numerical uncertainties is then obtained by subtracting in quadrature the experimental component from the total uncertainty, and is denoted with ``PDF+num''. In this analysis, the PDF and numerical uncertainties are strongly correlated, therefore their individual impacts are not estimated.

\begin{table}[htbp]
    \centering
    \begin{tabular}{c|cccc}
        \hline
        $\muk/2$ & \mtmukk & exp & PDF+num & scale \\
        {[GeV]} &[GeV] & [GeV] & [GeV] & [GeV]  \\ \hline
        
        192 & 160.90 & 0.66 & 0.80 & $+0.13, -0.69$ \\
        238 & 156.9 & 2.6 & 2.2 & $+1.4, -3.0$ \\ 
        322 & 152.9 & 4.5 & 4.2 & $+4.4, -6.7$ \\ 
        512 & 134.8 & 8.6 & 7.3 & $+9.0,-12.2$ \\ 
        \hline
    \end{tabular}
    \caption{Extracted values of \mtmukk and their uncertainties. The experimental (exp) component is obtained by freezing all the nuisance parameters to their post-fit values, while the combination between PDF and numerical uncertainties (PDF+num) is obtained by subtracting in quadrature the experimental component from the total uncertainty. The scale uncertainty refers to the variations of \mur and \muf.}
    \label{tab:masses}
\end{table}

\begin{figure}[htbp]
    \centering
    \includegraphics[width=.75\textwidth]{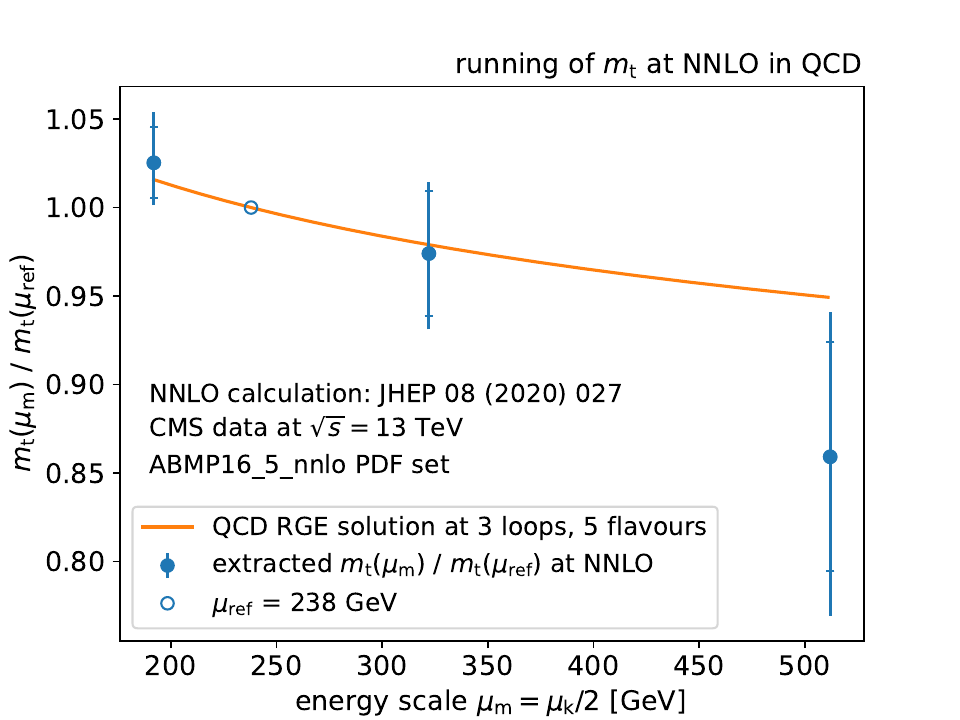}
    \caption{Extracted running of the top quark mass (full markers) normalised to the reference energy scale of 238~GeV (hollow marker), compared to the 3-loop solution of the QCD RGE assuming 5 active flavours (line). The inner vertical bars represent the combination of experimental, PDF, and numerical uncertainties, while the outer bars also include the QCD scale uncertainties.}
    \label{fig:running}
\end{figure}

Following the strategy of Ref.~\cite{CMS:2019jul}, the running is defined with respect to the reference scale $\mu_\mathrm{ref} = \mu_2/2 = 238~\mathrm{GeV}$. The choice of the reference scale is arbitrary, and does not affect the conclusions of the analysis. The quantities $r_k = \mtmukk/\mt(\mu_2/2)$ are derived and compared to the RGE prediction for $\mtmu/\mt(\mu_2/2)$. The advantage of this approach is the cancellation of the correlated components of the systematic uncertainties in the \mtmukk. Furthermore, the scale dependence of the QCD running is probed independently of the value of \mt. The RGE is solved at three loops in QCD assuming 5 active flavours, consistently with the calculation of Ref.~\cite{Catani:2020tko}, using the \textsc{CRunDec} program~\cite{Schmidt:2012az}. Good agreement between the measured points and the RGE prediction is observed, as shown in Figure~\ref{fig:running}. The reduced \chisq between the RGE and the measured $r_k$ is obtained in the Gaussian approximation by combining the covariance matrix from the \chisq fit to that corresponding to the scale variations. A reduced \chisq of 0.49 is obtained, which corresponds to a $p$-value of 69\%, reflecting the good agreement between the RGE prediction and the observed running of \mt. The compatibility with a hypothetical no-running scenario in which \mtmu is independent of \mum is also assessed, resulting in a reduced \chisq of 0.87 and a $p$-value of 46\%. An alternative scenario in which \mur and \muf scale variations are uncorrelated between the different \mtt bins is also considered, resulting in a reduced \chisq of 0.30 ($p$-value = 83\%) with respect to the RGE running, and of 1.0 ($p$-value = 39\%) for the hypothetical no-running scenario. Although this latter hypothesis cannot be excluded in this study, we conclude that the data indicate a clear preference for the RGE running scenario.

\section{Summary}
\label{sec:summary}

The running of the top quark mass is studied at next-to-next-to-leading order (NNLO) in quantum chromodynamics (QCD) for the first time. The analysis makes use of NNLO QCD predictions in the \msbar scheme based on the \Matrix framework~\cite{Grazzini:2017mhc} and implemented in Ref.~\cite{Catani:2020tko}, and of a differential measurement of the top quark-antiquark (\ttbar) production cross section from the CMS experiment at the CERN LHC~\cite{CMS:2019jul}. The running is extracted as a function of the invariant mass of the \ttbar system by means of a \chisq fit of the theoretical predictions to the measured cross section. The analysis benefits from a significantly improved fit procedure, developed for the purpose of this work, which consistently takes into account the numerical uncertainties in the calculation and their correlations with the other parameters of the fit. The extracted running is found to be in good agreement with the solution of the QCD renormalisation group equations (RGEs), within experimental and theoretical uncertainties. Although a hypothetical no-running scenario cannot be excluded, the result of this study indicates a clear preference for the RGE running hypothesis.

\acknowledgments

We are sincerely grateful to Massimiliano Grazzini,  Stefano Catani, Sven-Olaf Moch, and Andr\'{e} Hoang for fruitful discussions on the theoretical aspects of this work. J.M.\ is grateful to Stefano Catani, Simone Devoto, Massimiliano Grazzini, and Stefan Kallweit for their contribution in the development of the results of Ref.~\cite{Catani:2020tko}, and for many fruitful discussions. The work by K.L.\ is supported by the Helmholtz Association under the contract W2/W3-123.

\bibliographystyle{JHEP}
\bibliography{biblio.bib}

\end{document}